\documentclass[useAMS,usenatbib]{mn2e}
\usepackage{graphicx}
\usepackage{lscape}
\newcommand{\ha}{$H_\alpha$ }
\newcommand{\hb}{$H_\beta$ }
\newcommand{\todcor}{\textsc{todcor} }
\newcommand{\phoebe}{\textsc{phoebe} }
\newcommand{\jkt}{\textsc{jktebop} }
\newcommand{\yy}{Y$^2$ }
\title[A 0.61 + 0.45 M$_\odot$ detached eclipsing binary in a multiple system.]
{Orbital and physical parameters of eclipsing binaries from the ASAS 
catalogue --- IV. A 0.61 + 0.45 M$_\odot$ binary in a multiple system.}
\author[K. G. He{\l}miniak et al.]
{K. G. He{\l}miniak$^{1,2}$\thanks{E-mail:xysiek@astro.puc.cl},
M. Konacki$^{2,3}$, M. R\'{o}\.{z}yczka$^{4}$, J. Ka{\l}u\.zny$^{4}$, M. Ratajczak$^{2}$,
\newauthor
J. Borkowski$^{2}$, P. Sybilski$^{2}$, M. W. Muterspaugh$^{5,6}$, D. E. Reichart$^{7}$, K. M. Ivarsen$^{7}$, 
\newauthor
 J. B. Haislip$^{7}$, J. A. Crain$^{7}$, A. C. Foster$^{7}$, M. C. Nysewander$^{7}$, A. P. LaCluyze$^{7}$ \\
$^{1}$Departamento de Astronom\'{i}a y Astrof\'{i}sica, Facultad de F\'{i}sica,
Pontificia Universidad Cat\'{o}lica de Chile,\\ 
Av. Vicu\~{n}a Mackenna 4860,
782-0436 Macul, Santiago, Chile\\
$^{2}$Nicolaus Copernicus Astronomical Center, 
Department of Astrophysics, ul. Rabia\'{n}ska 8 , 
87-100 Toru\'{n}, Poland\\
$^{3}$Astronomical Observatory, A. Mickiewicz University, ul. S{\l}oneczna 36, 
60-286 Pozna\'{n}, Poland\\
$^{4}$Nicolaus Copernicus Astronomical Center, ul. Bartycka 18, 00-716 Warszawa, Poland\\
$^{5}$Department of Mathematics and Physics, College of Engeneering,
Tennessee State University, Boswell Science Hall,\\
Nashville, TN 37209, USA\\
$^{6}$Tennessee State University, Center of Excellence in Information
Systems, 3500 John A. Merritt Blvd., Box No. 9501,
Nashville, \\TN 37203-3401, USA\\
$^{7}$Department of Physics and Astronomy, University of North Carolina, 
Campus Boc 3255, Chapel Hill, NC 27599-3255\\
}

\begin{document}


\date{Accepted 2012 June 12.  Received 2012 June 8; in original form 2011 December 2}

\pagerange{\pageref{firstpage}--\pageref{lastpage}} \pubyear{2011}

\maketitle

\label{firstpage}

\begin{abstract}
We present the orbital and physical parameters of a newly discovered
low-mass detached eclipsing binary from the \emph{All-Sky Automated Survey}
(ASAS) database: {ASAS~J011328-3821.1~A} -- a member
of a visual binary system with the secondary component separated by about
1.4 seconds of arc. The radial velocities were calculated from the
high-resolution spectra obtained with the 1.9-m Radcliffe/GIRAFFE, 
3.9-m AAT/UCLES and 3.0-m Shane/HamSpec telescopes/spectrographs on the
basis of the \todcor technique and positions of \ha emission lines. For the 
analysis we used $V$ and $I$ band photometry obtained with the 1.0-m Elizabeth 
and robotic 0.41-m PROMPT telescopes, supplemented with the publicly available
ASAS light curve of the system.

We found that ASAS~J011328-3821.1~A is composed of two late-type dwarfs having
masses of $M_1 = 0.612 \pm 0.030$ M$_\odot$, $M_2 = 0.445 \pm 0.019$ M$_\odot$
and radii of $R_1 = 0.596 \pm 0.020$ R$_\odot$, $R_2 = 0.445 \pm 0.024$ R$_\odot$,
both show a substantial level of activity, which manifests in strong
\ha and $H_\beta$ emission and the presence of cool spots. The influence of the third 
light on the eclipsing pair properties was also evaluated and the photometric 
properties of the component B were derived. Comparison with several 
popular stellar evolution models shows that the system is on its main sequence
evolution stage and probably is more metal rich than the Sun. 
We also found several clues which suggest that the component B itself is
a binary composed of two nearly identical $\sim 0.5$ M$_\odot$ stars.
\end{abstract}

\begin{keywords}
binaries: eclipsing -- binaries: spectroscopic -- binaries: visual -- 
stars: fundamental parameters -- stars: individual (ASAS~J010538-3821.1) 
-- stars: low-mass
\end{keywords}

\section{Introduction}

In the last decade there has been a stunning increase in our knowledge 
of the structure and evolution of 
eclipsing binaries. The number of known and well studied 
detached low-mass eclipsing systems, which are the main source of absolute 
parameter measurements, increased by almost an order of magnitude after the 
year 2000. This increase helped several authors make 
conclusions about a 30-year old problem: the discrepancy between 
observed and theoretically predicted radii and temperatures during the 
evolution of low-mass stars \citep{lac77}.
The prevailing idea is that this discrepancy is caused by the stellar
magnetic field, amplified by the fast rotation due to tidal locking in close 
binaries, which inhibits the efficiency of convection in the envelope and 
manifests itself in a higher level of activity than for single stars \citep{cha07}. 
However, the number of known well-characterized systems is too low to 
reliably test this hypothesis. This is especially true amoung the long-period 
systems, which normally exhibit lower levels of activity, so the observed 
radii and temperatures are closer to the theoretically predicted values. 
In a very recent paper, \citet{irw11} 
described an eclipsing pair of M dwarfs on a 41 d orbit which, despite slow
asynchronous rotation and a probable age over 120 Myr, still show cold spots,
and radii are inflated by 5\% with respect to theoretical predictions. 
This system is evidence against the theory mentioned above and may again 
bring wider interest to the need to revisit the equation of state for low mass
stars, which was suggested to explain the inflated radii before the activity
hypothesis \citep[e.g.][]{tor02}.

To be useful for testing the evolutionary models, the parameters of the system,
especially masses and radii, should be known with a precision of at least 3\%, 
considered now as a canonical level \citep{bla08,cla08}. 
There are two other criteria which can make a system more interesting: 
(1) additional multiplicity -- by deriving properties of the third body one 
can put additional constraints on the nature of the whole system; 
(2) low mass ratio -- in general it is harder to fit a single isochrone to 
data points that are well separated in parameter space; these are usually 
stars of different masses. Twin stars commonly have data points that lie close 
together and are less difficult to fit. In this paper we present our 
analysis together with orbital and physical parameters of a newly discovered
low-mass detached eclipsing binary which meets the two mentioned criteria.
The data we gathered fell short of the 3\% precision level goal, but the 
system is still highly valuable for our understanding of the nature of 
low-mass stars.

\section{The ASAS J011328-3821.1 system}

The eclipsing system ASAS~J011328-3821.1 (2MASS J01132817-3821024,
1RXS~J011328.8-382059; hereafter ASAS-01) has the shortest orbital period
from all objects in our sample of low-mass detached eclipsing binaries
\citep[LMDEBs; see:][]{hel11a,hel11b} found in the \emph{ASAS Catalogue of Variable Stars} 
\citep[ACVS;][]{poj02}\footnote{\texttt{http://www.astrouw.edu.pl/asas/?page=acvs}},
and one of the shortest among the LMDEBs known to date -- 
$P_{ASAS} = 0.44559$ d. In the ACVS the ASAS-01 is classified
as eclipsing detached (ED) with a maximum brightness of $V = 11.78$ mag
and amplitude of brightness variation of 0.33 mag. The ASAS light curve
clearly shows that there is a significant difference in the eclipse depths which
implies that the eclipsing pair is composed of two vastly different stars.

Before the ACVS the object was not reported to be a binary or multiple but 
was noted as an X-ray source in the \emph{Einstein} \citep{gio90,sto91} 
and ROSAT \citep{vog99} catalogues. Later \citet{bee96} noted emission 
in calcium H and K lines, which they described in their catalogue as 
``moderate to strong''. More recently \citet{ria06} identified ASAS-01 as an 
M0.5 dwarf, gave values of spectral indexes CaH1, CaH2, CaH3 and TiO5, 
reported \ha emission line with equivalent width of 2.5 \AA, and estimated 
the distance to be about 37 pc. This determination is significantly different 
from the value of 26.6 pc obtained by \citet{szc08}, who derived
it on the basis of a bolometric correction derived from the ASAS $V-I$
colour, but the object was treated as a single star, explaining the discrepancy.

The emission in \ha was recently confirmed by \citet{par09} who
estimated the spectral types of components to be M1V + M3V. Their study, 
however, did not reflect the possibility of the system being composed of 
3 or more stars. 

Finally, in their lucky-imaging survey of southern M dwarfs, \citet{ber10} 
obtained high-angular-resolution images of the system in $i'$ and $z'$ bands. 
They clearly separated two components, with the secondary located 1.405(3)
seconds of arc at the position angle of 29.0(3) degrees.
They estimated the spectral types to be M0 + M1, not taking into 
account the eclipsing nature of one of the stars.

\begin{figure}
\includegraphics[width=\columnwidth]{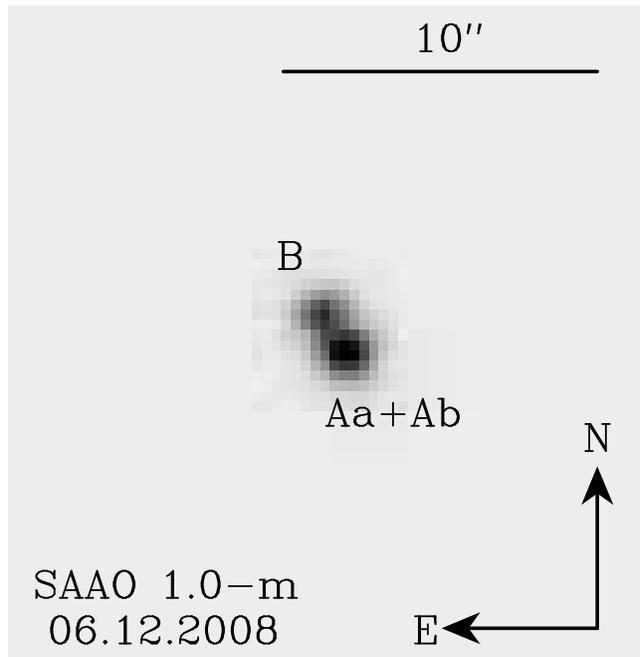}
\caption{A part of an image taken with 1.0-m Elizabeth telescope in SAAO
showing the two components of the ASAS-01~AB system. The visible components
are separated by $\sim$1.25 asec. ASAS-01~Aa~+~Ab is a low-mass eclipsing 
detached binary identified in the ASAS database ($P_{ASAS} = 044559$ d). 
The orientation and the scale of the image are shown.
}\label{fig_foto}
\end{figure}

We deduced the multiple nature of ASAS-01 in two ways. Firstly we noted that in 
the spectrum taken with the Shane telescope in 2007 (Sect. \ref{sec_obs_sp}) 
the \ha emission line profile was found to be triple. The same feature was found 
in a spectrum obtained in 2008 with the Anglo-Australian Telescope but not in
data obtained in 2006 with the Radcliffe telescope in South Africa. 
Shortly thereafter and independent of \citet{ber10} we found the two components 
clearly separated on the CCD images obtained with the Elizabeth telescope 
in South Africa (Sect. \ref{sec_obs_ph}) in good seeing conditions. 
A part of our image is shown in Fig. \ref{fig_foto}. The fainter B component 
is located about 1.25 arcseconds from the brighter A in the NE direction, 
which is consistent with the more precise result obtained by \citet{ber10}. 
A series of images taken in different orbital phases (with respect to the 
photometric 0.44559 d period) allowed us to deduce that eclipses occur in 
the component A.

\section{Spectroscopy and radial velocities}\label{sec_obs_sp}

\subsection{Observations}
To directly derive masses of ASAS-01 components we obtained a series
of high-resolution spectra in order to measure their radial velocities (RV).
Most of the spectroscopic 
observations come from the 1.9-m Radcliffe telescope and its {\it Grating 
Instrument for Radiation Analysis with a Fiber-Fed Echelle} (GIRAFFE) 
spectrograph at the South African Astronomical Observatory (SAAO), with 
settings the same as described in previous papers of this series 
\citep{hel09,hel11b}. Five spectra of $R\sim40000$,
were obtained during three consecutive nights in September 2006. After
3600 s of exposure and binning to $R\sim10000$ we obtained $SNR$ at $\lambda = 6520$
\AA \,between 25 and 40. At that time we were not aware that the system is 
also a visual binary, and the quality of the image from the aquisition camera 
was not enough to demonstrate that. Due to a relatively high seeing (about 2 asec 
and more) our spectra were contaminated by the companion B.

One additional spectrum was obtained
in October 2007 with the 3.0-m Shane telescope and its {\it Hamilton Spectrograph} 
(HamSpec) at the Lick Observatory. The target was observed very low 
over the horizon for 2100 seconds. After binning from the original resolution
of 60000 to 30000 we reached the $SNR$ level around 25 at $\lambda = 6560$ \AA.
Due to the low $SNR$, fringing, telluric lines and limited extent of available 
templates we did not use all 96 rows of the spectrum, but only used 21 that 
covered the wavelength range from 5150 to 6500~\AA. This spectrum was also 
contaminated by the light from the component B.

The last spectrum was obtained in September 2008 with the 3.9-m Anglo-Australian 
Telescope and its {\it University College London Echelle Spectrograph} (UCLES) 
at the Siding Spring Observatory, with the same settings as described in 
\citet{hel09,hel11b}. This spectrum was not binned and after 900 s
of exposure we reached $SNR\sim20$ at $\lambda = 6530$ \AA. We tried to
keep the slit on component A only, by rotating it by $90$ deg from the 
position angle of the AB pair, but due to the high seeing (over 2 asec) 
we could not avoid a significant contamination from star B.

\subsection{Radial velocities derivation.}
The CCD reduction, spectrum extraction and wavelength calibration were done
in every case with standard \textsc{iraf} procedures\footnote{\textsc{iraf} 
is written and supported by the \textsc{iraf} programming group at the National 
Optical Astronomy Observatories (NOAO) in Tucson, AZ. NOAO is 
operated by the Association of Universities for Research in 
Astronomy (AURA), Inc. under cooperative agreement with the 
National Science Foundation. \texttt{http://iraf.noao.edu/}}. For the
wavelength calibration we used exposures of ThAr lamps taken before and
after the science exposure. Initially the RV's were measured with our
implementation of the two dimensional cross-correlation technique
\citep[\textsc{todcor};][]{zuc94} using a spectrum of an M2V standard 
star HIP~93101 broadened to $v_{rot} \sin{i}=30$~km~s$^{-1}$
that was taken as a template. The formal RV measurement errors were computed 
from the bootstrap analysis of \todcor maps created by adding randomly 
selected single-order \todcor maps.

In the \todcor maps, the location of the maximum of 
the cross-correlation function in the $v_1/v_2$ plane refers to radial 
velocities of the two components of the system. We could easily phase-fold 
our measurements of the primary $v_1$ with the photometric period of 
0.44559 d, but we failed when trying to do this for $v_2$ (putative secondary). 
Thus to obtain the RV curve of the eclipsing secondary we decided to measure 
positions of the \ha emission lines. In all GIRAFFE
spectra we could easily recognize two well-separated components of that line and 
measure their positions despite that the line was close to the edge of an 
\'{e}chelle order. To measure the positions we used the \texttt{deblend} procedure
in the \textsc{iraf/echelle} package and fitted a double-gaussian function. The fit
was performed about 20 times with different initial wavelength range selections,
initial peak positions and smoothing factors. Results were then averaged and 
the standard deviation was taken as the measurement error. A similar procedure
was applied to the HamSpec and UCLES spectra but with a triple-gaussian
function, since the triple character of the \ha line was obvious. We did
not normalize the spectra to the continuum before this procedure.

To check whether the RV's obtained from \ha lines can be trustful we tried 
to phase-fold them separately with the photometric period. For the 
primary we found a good agreement with the \todcor measurements, and for 
the secondary we succeeded to find a fit with the desired period, systemic 
velocity close to the one found for the primary, and amplitude $K_2$ 
significantly higher than the primary's $K_1$. This is expected 
for two distinctive late-type dwarfs, considering the difference in the 
depths of the eclipses in the ASAS light curve. For the final orbital fit we 
used all measurements simultaneously. We used the spectroscopic orbit 
fitting procedure described in our previous papers \citep{hel09,hel11a,hel11b}. 
This simple code uses a Levenberg-Marquardt minimalization algorithm to find a 
keplerian orbit of a spectroscopic binary. It also allows for estimation of 
systematic contributions to the error budget by Monte-Carlo and bootstrap 
analysis.

\begin{table}
\centering
\caption{Single RV measurements of all components of the ASAS-01 
system derived with the \todcor technique and from the position 
of \ha emission lines, together with their formal errors and
final orbital fit residua. ``R/G'' denotes data points from
Radcliffe/GIRAFFE observations, ``S/H'' from the Shane/HamSpec
spectrum and ``A/U'' from AAT/UCLES. For the B component the 
residua refer to the fit of a circular orbit with 
$P_B \simeq 0.475$~d. When $O-C$ is not given, the measurement
was not included into the fit.}
\label{tab_rv}
\begin{tabular}{lrrrc}
\hline \hline
JD & $v$ & $\pm$ & $O-C$ & T/S \\
$-2450000$& (km s$^{-1}$) & (km s$^{-1}$) & (km s$^{-1}$) & \\
\hline
\multicolumn{5}{l}{\it Component Aa velocities from \todcor}\\
 4006.50332 & -86.882 & 4.360 &  3.502 & R/G \\
 4007.52761 &  98.787 & 7.240 &  3.361 & R/G \\
 4007.57128 & 140.808 & 7.037 &  7.444 & R/G \\
 4008.47697 & 140.128 & 4.695 &  3.230 & R/G \\
 4008.52042 & 112.337 & 4.540 & -6.065 & R/G \\
 4375.87729 &-100.752 & 8.175 & -1.146 & S/H \\
 4727.14578 &  68.250 & 1.247 &  0.076 & A/U \\
\multicolumn{5}{l}{\it Component Aa velocities from \ha lines}\\
 4006.50332 & -90.330 & 6.971 &  0.054 & R/G \\
 4007.52761 &  96.905 & 7.435 &  1.479 & R/G \\
 4007.57128 & 132.301 & 6.895 & -1.063 & R/G \\
 4008.47697 & 122.606 & 7.232 &  ---   & R/G \\
 4008.52042 & 107.616 & 6.577 &-10.786 & R/G \\
 4375.87729 &-110.900 & 6.563 &-11.295 & S/H \\
 4727.14578 &  94.415 & 6.540 &  ---   & A/U \\
\multicolumn{5}{l}{\it Component Ab velocities from \ha lines}\\
 4006.50332 & 161.590 & 7.509 & -7.093 & R/G \\
 4007.52761 & -85.570 & 7.401 &  1.556 & R/G \\
 4007.57128 &-144.540 & 7.134 & -5.276 & R/G \\
 4008.47271 &-136.700 & 7.866 &  6.691 & R/G \\
 4008.52042 &-113.824 & 7.176 &  4.881 & R/G \\
 4375.87729 & 191.210 & 6.655 &  9.812 & S/H \\
 4727.14578 & -68.139 & 6.525 &  ---   & A/U \\
\multicolumn{5}{l}{\it Component B velocities from \todcor}\\
 4006.50332 & 116.254 & 6.240 &  3.502 & R/G \\
 4007.52761 &  28.122 & 1.500 &  3.361 & R/G \\
 4007.57128 & -21.130 & 6.076 &  7.444 & R/G \\
 4008.47697 &  19.200 & 3.690 &  3.230 & R/G \\
 4008.52042 & -24.688 & 9.086 & -6.065 & R/G \\
 4375.87729 & -12.785 & 5.925 & -1.147 & S/H \\
 4727.14578 &  21.381 & 0.482 &  0.076 & A/U \\
\multicolumn{5}{l}{\it Component B velocities from \ha lines}\\
 4375.87729 &  13.147 & 6.587 &  ---   & S/H \\
 4727.14578 &	6.799 & 6.633 &  ---   & A/U \\
\hline
\end{tabular}
\end{table}

\subsection{ASAS-01 A orbital fit}

\begin{figure*}
\includegraphics[width=0.9\textwidth]{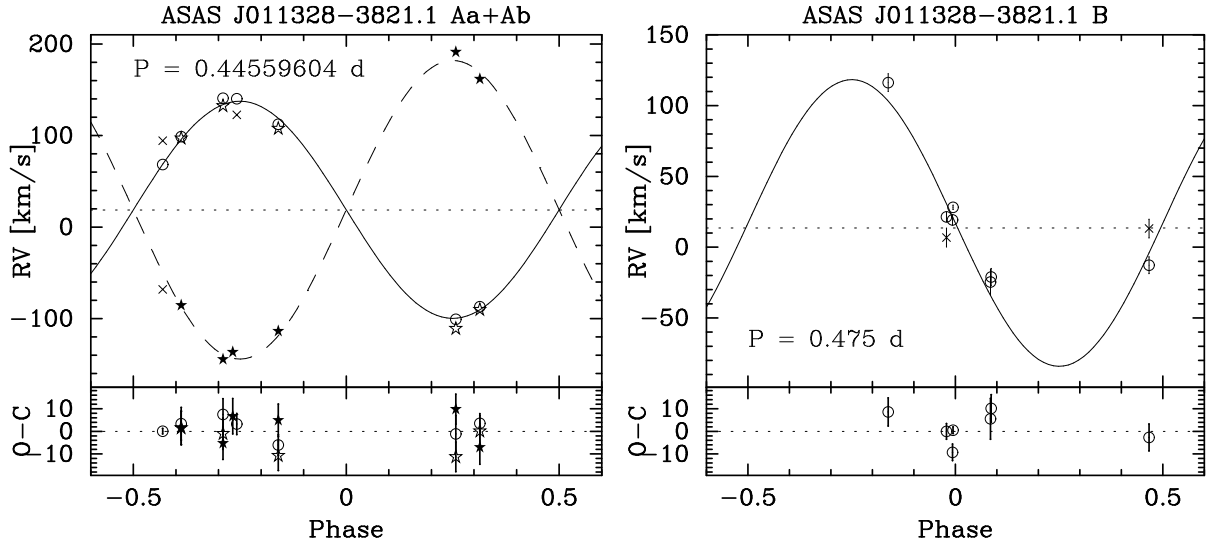}
\caption{{\it Left:} Radial velocity measurements and the best-fitting 
orbital solution for the ASAS-01~A eclipsing binary. Open symbols and
the solid line are for the primary, filled symbols and the dashed line are
for the secondary. Circles denote \todcor, stars denote \ha lines 
measurements and crosses mark measurements rejected from the final fit.
Residua with the final error bars are shown on the lower panel;
{\it Right:} Radial velocity measurements of the ASAS-01~B component
from the \todcor technique and the best fit of a putative circular orbit
of $P\simeq0.475$ d period. Crosses mark measurements rejected from the final fit.
Residua and error bars are shown on the lower panel.
}\label{fig_rv}
\end{figure*}

Before obtaining the orbital fit we used our photometric measurements (Sect.
\ref{sec_obs_ph}) to correct the ASAS ephemeris. Using the \jkt 
and \phoebe codes (see Sect. \ref{sect_model}) we found a new value of the 
orbital period $P = 0.44559604(18)$~d and incorporated it to the orbital
fit. We have also found (from the photometry and radial velocities) 
that the eccentricity $e$ agrees with zero well
within formal errors, so we held it fixed later. We found the 
same for the difference between systemic velocities of the primary and secondary 
$\gamma_{Aa} - \gamma_{Ab}$, and for any possible systematic shifts 
between $v_1$ measurements from \todcor and \ha lines. However, the 
reduced $\chi^2$ of the best-fitting solutions made separately for \ha 
measurements of either the primary or the secondary were not close to 1. 
This means that the measurement errors were underestimated, so we
added in quadrature a systematic term of 6.5 km~s$^{-1}$ to the formally 
derived measurement errors. This operation was not necessary for the 
\todcor results. We also excluded one \ha measurement of the primary from the
GIRAFFE spectrum at JD$\sim$2454008.477 and both from the UCLES 
spectrum, due to their very large formal uncertainties. The measurements 
from the UCLES spectrum probably also suffered from the line blending. For this 
exposure the velocity difference was the smallest in the whole sample and the
emission from the B component is strongest. However, we believe that the
\todcor measurement is secure because the peak was very high and well separated 
from other smaller maxima.

All the RV mesurements for all components of the ASAS-01 multiple system are listed
in Table \ref{tab_rv}. For ASAS-01~A we show only the measurements which were
taken to produce the orbital fit and the final model of the system, together with their
final errors and residuals. For the component B we show all the \todcor $v_2$ 
values and two \ha velocities derived from HamSpec and UCLES spectra. In the 
``T/S'' column ``R/G'' denotes Radcliffe/GIRAFFE, ``S/H'' denotes 
Shane/HamSpec and ``A/U'' denotes AAT/UCLES telecope/spectrograph.

\begin{table}
\caption{Orbital parameters of the ASAS-01~A eclipsing binary.}
\label{tab_orb}
\centering
\begin{tabular}{l r l}
\hline \hline
Parameter & Value & $\pm$ \\
\hline
$P_A$ [d] 		& 0.44559604 & 1.8e-7 \\
$T_{0,A}$ [JD-2450000]	& 1868.83933 & 3.1e-4 \\
$K_{Aa}$ [km~s$^{-1}$] 	& 118.4 & 2.0 \\
$K_{Ab}$ [km~s$^{-1}$] 	& 162.9 & 3.3 \\
$\gamma_A$ [km~s$^{-1}$]&  18.6 & 2.3 \\
$q_A$ 		 	& 0.727 & 0.018 \\
$a_A \sin{i}$ [R$_\odot]$ & 2.478 & 0.034 \\
$e_A$ 		 	&  0.0  & (fix) \\
$M_{Aa} \sin^3{i}$ [M$_\odot]$ & 0.595 & 0.027 \\
$M_{Ab} \sin^3{i}$ [M$_\odot]$ & 0.432 & 0.017 \\
\hline
\end{tabular}
\flushleft
\scriptsize
Note: Values of $P$, $T_0$ are 
taken from the simultaneous spectroscopic and photometric analysis 
performed with \textsc{phoebe}.
\end{table}

In the left panel of Fig. \ref{fig_rv} we present all the measurements together
with the best fitting orbital solution obtained for the ASAS-01~A eclipsing pair.
We found the solution with a fast procedure which fits a double-keplerian orbit
by minimizing the $\chi^2$ function with a Levenberg-Marquard method.
The $rms$ of the fit was 5.8 km~s$^{-1}$ for the primary and 7.0 km~s$^{-1}$ for 
the secondary. The final reduced $\chi^2$ was 0.988, thus we believe that the 
parameter uncertainties are well estimated. To ensure the accuracy of the 
estimates we also performed an additional Monte-Carlo simulation to estimate 
the systematic errors of derived parameters. The resulting systematic terms
were an order of magnitude smaller than the formal ones, but
were added in quadrature to the errors from the least-squares fit.

The orbital parameters derived from the fit are listed in Table \ref{tab_orb}.
One can see that \ha and \todcor measurements are complementary and, when combined,
allow us to reach about 2\% level of precision in the velocity amplitudes $K_{1,2}$
and about 4\% in $M\sin^3{i}$. We would like to emphasize the low value of the
mass ratio $q = 0.727(18)$, which makes ASAS-01~A quite unique. With this $q$
ASAS-01~A resides in the mass-ratio distribution of spectroscopic binaries
well outside both the narrow peak at $q \ga 0.95$ \citep[the ``strong'' 
twin hypothesis;][]{luc06} and the wide one at $q \ga 0.85$ 
\citep[the ``weak'' twin hypothesis;][]{hal03}. This makes ASAS-01~A
particularly interesting for testing the evolutionary models.

\subsection{The component B}
The positions of the third \ha emission peak in the HamSpec and UCLES 
spectra were close to (however not in good agreement with) 
the corresponding measurements of $v_2$ from \todcor. This led to a 
conclusion that the $v_2$ coordinate of the \todcor maxima may be related 
to the B component, not the secondary component of the eclipsing Aa+Ab pair. 
If so, the B component itself must be a spectroscopic binary, since the $v_2$ 
values change significantly by about 150 km s$^{-1}$. The third emission \ha 
peak is however not clearly visible in the GIRAFFE spectra so we can't fully 
support that hypothesis by comparing all \todcor and \ha measurements. 
Nevertheless it is worth noticing that during two consecutive nights when
two R/G observations were taken with roughly one hour time span 
(JD$\sim$2454007.5 and 2454008.5) the position of the peak on the \todcor map
changed by about 45 km~s$^{-1}$. The remaining R/G spectrum
(JD$\sim$2454006.5) -- from which the highest value of $v_2$ was derived
(116~km~s$^{-1}$) -- had the highest signal to noise ratio.

\begin{table}
\caption{Parameters of the ASAS-01~B putative orbit.}
\label{tab_rv_B}
\centering
\begin{tabular}{l r l}
\hline \hline
Parameter & Value & $\pm$ \\
\hline
$P_B$ [d] 		& 0.4750 & 0.0009 \\
$T_{0,B}$ [JD-2450000]	& 4806.950$^a$ & 0.022 \\
$K_{Ba}$ [km~s$^{-1}$] 	& 104.7 & 5.7 \\
$\gamma_B$ [km~s$^{-1}$]& 13.6 & 2.9 \\
$e_B$ 		 	&  0.0  & (fix) \\
$a_{Ba} \sin{i}$ [R$_\odot]$ & 0.983 & 0.053 \\
$f(M_{Ba})$ [M$_\odot]$ & 0.0565 & 0.0092 \\
\hline
\end{tabular}
\flushleft
\scriptsize
$^a$ $T_0$ is calculated for the moment of the orbital conjunction.
\end{table}

From the \todcor measurements of the B component's RVs we derived
an acceptable orbital solution with the $rms$ of measurements of 7.1~km~s$^{-1}$
and reduced $\chi^2$ of the fit nearly 3. The parameters of this solution
are listed in Table \ref{tab_rv_B} and the model RV curve, together with the
single measurements and residua (listed in Table \ref{tab_rv}), 
are plotted on the right panel of the Fig. \ref{fig_rv}. In this orbital fit 
we did not include the measurements derived from the \ha lines, because
the resulting parameters were not very different from the ones given, and the
quality of the fit (in terms of $rms$ and $\chi^2$) was substantially worse.
One can see that the systemic velocity $\gamma_B$ differs from the 
$\gamma_A$ value by less than the sum of their errorbars, which is expected
in case of gravitational bounding. This is an 
argument for the correctness of this fit and thus the binary nature of 
ASAS-01~B, but considering the amount and quality of the available data,
we are far from making any conclusive statements about this
hypothesis. The binarity of ASAS-01~B will be discussed in a later part
of the paper.

\subsection{\ha line profiles}

\begin{figure}
\centering
\includegraphics[width=\columnwidth]{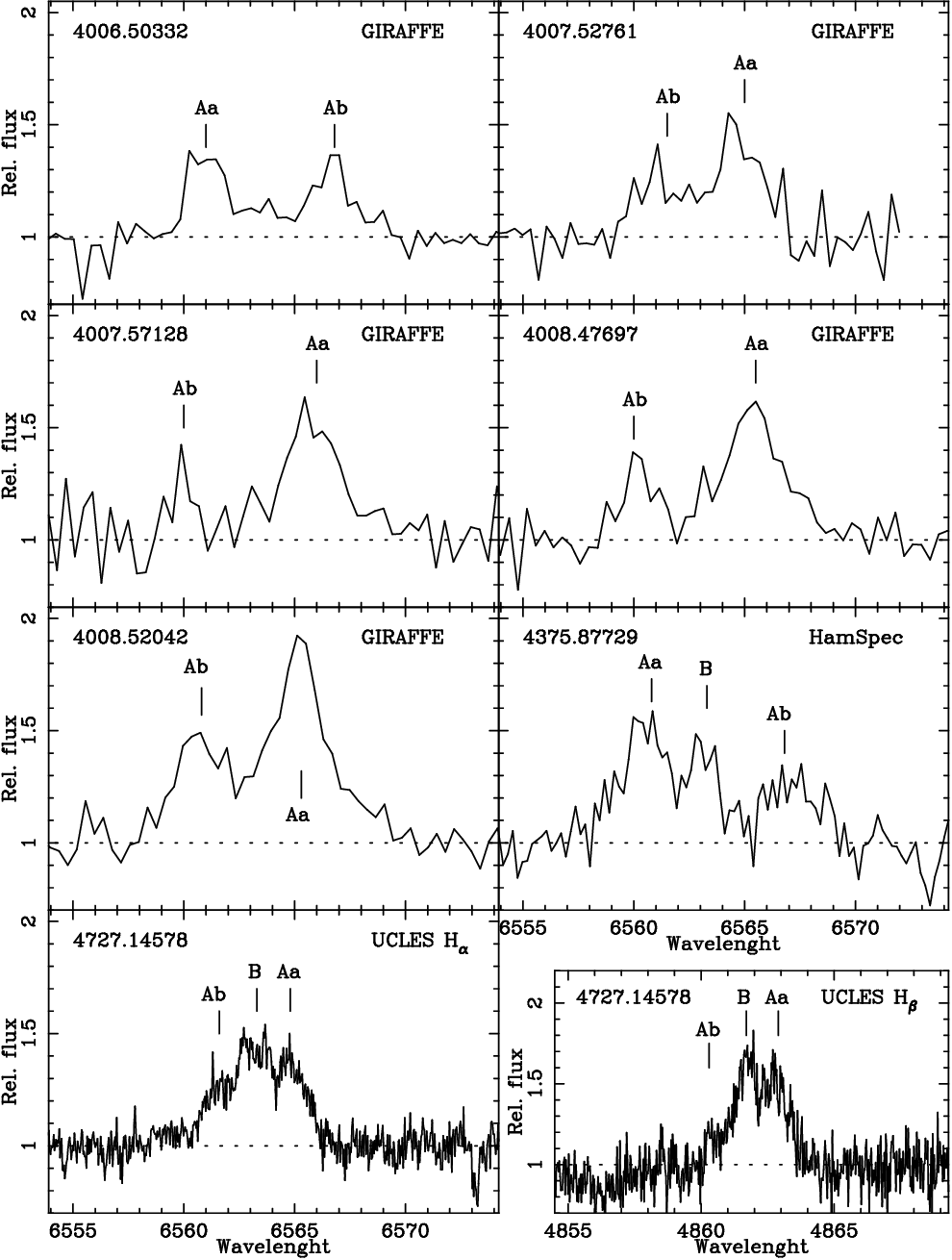}
\caption{ \ha emission profiles variability as seen in our
spectra, especially for the B component. For the GIRAFFE setup the
\ha line is close to the order's edge, so the $SNR$ around it is very
poor, which makes the continuum correction and $EW$ calculations
quite challenging. The \hb line from the 
UCLES spectrum is also shown for comparison. JD-2450000 and spectrographs are
labelled, components are marked. Spectra were continuum normalized.
}\label{fig_ha}
\end{figure}

The existence of strong \ha emission is a manifestation 
of a substantial stellar activity in all components
of the ASAS-01 system. The closer inspection of the \ha lines reveals their
probable variability. In Fig. \ref{fig_ha} we present \ha
profiles from all of our spectra. On the spectra from HamSpec and 
UCLES all three components are clearly 
visible and marked.
One has to remember that for the GIRAFFE setting the \ha line was close to 
the edge of the echelle order, so the $SNR$ around the line was lower.
For every spectrum we measured the \ha equivalent
width and collected them in Table \ref{tab_ha_ew}.

\begin{table}
\centering
\caption{\ha lines equivalent width measurements for all our spectra.
The measurement errors are estimated to be around 0.1\AA. Values for the B
component from the GIRAFFE spectra (R/G)
should be treated with caution.}\label{tab_ha_ew}
\begin{tabular}{lcccc}
\hline \hline
JD 	& $EW_{Aa}$ & $EW_{Ab}$ & $EW_{B}$ & T/S \\
-2450000& [\AA] &  [\AA] & [\AA] & \\
\hline
4006.50332& 0.74 & 0.54 & 0.18 & R/G \\
4007.52761& 1.38 & 0.71 & 0.15 & R/G \\
4007.57128& 1.39 & 0.40 & 0.13 & R/G \\
4008.47697& 1.69 & 0.65 & 0.10 & R/G \\
4008.52042& 2.04 & 0.58 & $<0.1$ & R/G \\
4375.87729& 1.60 & 0.67 & 0.60 & S/H \\
4727.14578& 0.60 & 0.25 & 0.87 & A/U \\
\hline
\end{tabular}
\end{table}

The Aa component seems to be particularly strong on both GIRAFFE spectra
from Sept. 30 (JD$\sim$2454008.5). This may be due eithe to the real increase 
of the line intensity or to blending with the B component. However,
in the latter case the velocity of B calculated from the line position
would differ by 50-70 km~s$^{-1}$ from the maximum on the coresponding
\todcor map. In the case of GIRAFFE data the intensity and even the 
existence of the third peak is highly dependent on how the spectrum was 
normalized to the continuum.
The low $SNR$ of the GIRAFFE spectra and the high rotational broadening of
the stellar lines make this operation quite challenging and its results
unreliable, especially at the orders' edges (i.e. around the \ha line).

The $EW$ measurements, given in the Table \ref{tab_ha_ew}, were 
made under assumption of three emission components for all spectra, 
but we stress that for R/G observation this assumption may be wrong
due to low SNR. This data set however shows a systematic decrease of 
the $EW$ measurements for the B component (by a factor of $\sim2$), 
and a rising trend for the Aa (by a factor of nearly 3). The $EW_{Ab}$ values 
seem to be randomly oscillating around a constant value. Nevertheless 
these measurements are unsecure and should be treated with caution.

In the HamSpec and UCLES observations the B component is obviously stronger
than in those performed with GIRAFFE. The Aa component shows values 
substantially different between the two spectra. The $EW_{Ab}$ for the 
HamSpec spectrum is within the oscillations seen in the GIRAFFE data, but in 
the UCLES spectrum it reaches the smallest value of all data sets. 
This may however be due to the blending with the component B. Given the 
activity of the system, we expect the observed variability of \ha to be 
real, however the quality of the data does not allow for a firm statement.

We also want to note that in three of five GIRAFFE spectra, as well as
in HamSpec and UCLES observations, we have found signatures of \hb emission 
lines. Especially in the UCLES spectrum the emission from all three components
is clear and the $EW$ measurements were possible. The shape of the
profile is very similar to the \ha line. We found $EW$'s to be 0.70, 0.17 and 
0.62 \AA\, for the Aa, Ab and B components respectively, with the 
uncertainty of $\sim$0.1 \AA. The \hb line from the UCLES spectrum is also
shown in Fig. \ref{fig_ha}.

\section{Photometric observations}\label{sec_obs_ph}

The photometric observations of ASAS-01 were carried out in $V$ and $I$ 
bands with the 1.0-m Elizabeth telescope, located in SAAO, and the 
0.41-m Prompt-4 and Prompt-5 robotic telescopes\footnote{\textit{Panchromatic 
Robotic Optical Monitoring and Polarimetry Telescopes}.
PROMPT is operated by SKYNET -- a distributed 
network of robotic telescopes located around the world, dedicated for 
continues GRB afterglows observations. \texttt{http://skynet.unc.edu}}, 
located in the Cerro Tololo Inter-American 
Observatory in Chile. A more detailed description of the observational
settings, reduction procedure and calibration to standard photometric
system can be found in \citet{hel11a} and \citet{hel11b}.

The majority of the SAAO observations were done in December 2008. Most of the 
orbital phases were covered, however without the primary minimum in $V$
and with only few points in $I$. A clear out-of-eclipse modulation, 
probably originated by a presence of a cold spot, was noticed.

The PROMPT observations were carried out mostly in July 2009, with two 
additional nights at the ends of August and September the same year. This time
several primary minima were covered, however more points were collected around 
the secondary minimum. Again we noticed an out-of-eclipse modulation which 
was bascially the same as for the former SAAO data in 
regards to its shape.

The last small part of the data, mostly in $V$, comes from SAAO observations 
conducted in October 2009. Again a wide range of the orbital phases was covered, 
including both minima, and an out-of-eclipse modulation was noticed to be of the 
same shape as for previous data sets. This led to a conclusion that if the
modulation comes from the cold spots, the spot pattern did not change much
between December 2008 and October 2009.

In total we gathered 598 brightness measurements in $V$ and 586 in $I$,
some under quite challenging conditions, which
gave us a full phase coverage. Several primary and secondary minima
as well as an out-of-eclipse variation were recorded,
which allowed us to perform a detailed modeling of the system, starting from
an update of the ephemeris and ending with spot and third light properties. 
For the last purpose especially useful were SAAO observations from
December 2008, where in good seeing conditions we were able to distinguish
A and B components on the CCD images (Fig. \ref{fig_foto}). From a number
of images where the components were well detached, we estimated their
brightness ratio in both bands. In this way we deduced that eclipses occur
in the brighter component A, and obtained reasonable starting values
for the third-light contribution, which were later incorporated in the model.

\section{The physical model}\label{sect_model}

\begin{figure*}
\includegraphics[width=0.92\textwidth]{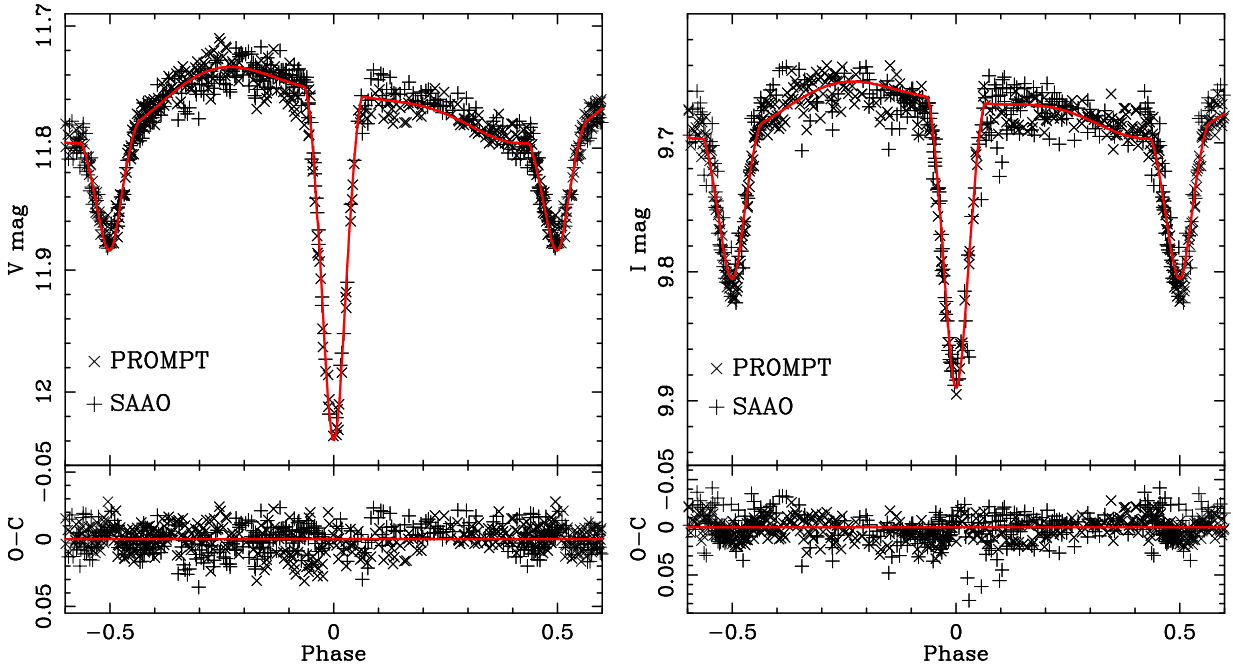}
\caption{The observed $V$ (left) and $I$ band (right) light curves of ASAS-01 
obtained from our SAAO ($+$) and PROMPT ($\times$) observations. An overplotted
red line shows the best-fitting model. An out-of-eclipse light modulation,
coming from the presence of two cold spots, is clearly visible. A significant
change of the difference between the depths of the minima in $V$ and $I$ bands 
indicate the temperature ratio is substantially different from unity. Residua are
plotted on the lower panels. Colour figure is available in the on-line version
of the manuscript.
}\label{fig_lc}
\end{figure*}

The modeling procedure was in many aspects identical to the ones described 
in the three previous papers of this series. We used the \jkt code
\citep{sou04a,sou04b}, which is based on the \textit{Eclipsing Binaries 
Orbit Program} \citep[\textsc{ebop};][]{pop81,etz81}, and \phoebe 
\citep{prs05} -- an implementation of the \textsc{wd} code \citep{wil71}. 
The \jkt was used on the ASAS light curve and served mainly to correct 
the ASAS ephemeris and check for marks of a non-zero eccentricity. We did not 
use it extensively since it does not allow for spots or work on RV 
curves. We should also note that PROMPT and SAAO light curves are almost 
indistinguishable and initial fits performed separately gave similar results, 
thus we decided to use all photometric data sets simultaneously in the whole 
fitting process. Photometric data points were weighted
by their formal errors. Absolute values of several physical parameters, like
radii, bolomeric magnitudes and distance, were obtained with the 
\textsc{jktabsdim} code, available with \textsc{jktebop}, in the same way
as decribed in previous papers.

We improved the ephemeris from \jkt with \phoebe using all available data, 
and used them to create a final orbital fit (Sect. \ref{sec_obs_sp}) and later
the full solution. We set the third light level in both bands to the values 
obtained from the CCD images and let them be fitted for in later stages of 
the process. We consider the uncertainty in the third light level as the 
main source of errors of physical parameters such as radii, temperatures and
component magnitudes in both bands \citep[see also ASAS-08 in:][]{hel11a}. 
The third light uncertainty may possibly have its origin in the intrinsic 
photometric variability of component B, which is expected for such 
an active star.

The starting values of effective temperatures
were deduced from the empirical colour--temperature relations by \citet{wor10}
on the basis of dereddened $V-I$ values for every component. For the 
dereddening we used $E(B-V)=0.0145$ value from \citet{sch98}. In general 
\phoebe allows for calculating fluxes in every passband separately for 
every star. If one has at least two light curves in various bands, the program 
also allows for fitting both effective temperatures simultaneously using the 
concept of a ``binary effective temperature'' \citep{prs05}. The treatment of limb 
darkening and reflection albedos was the same as in our previous papers, as well as
the way of finding parameters of cold spots. We tested several different configurations
and found the best solution is when a large cold spot is located on the secondary and
is obscured during the secondary eclipse, and a second small spot with maximum
appearance around phase $\phi=0.25$ on the primary. Because the light curves do not
contain much information about the latitude of spots, we decided to locate spots
near the stars' equators and keep this location fixed. Moving towards other latitudes did 
not change the final model significantly -- the difference in inferred values of 
stellar parameters was smaller than uncertainties coming from other sources.
 
\begin{table}
\centering
\caption{Absolute physical parameters of the ASAS-01~A eclipsing pair.}\label{tab_par}
\begin{tabular}{lrlrl}
\hline \hline
{\bf ASAS-01~A} & \multicolumn{2}{c}{Primary}& \multicolumn{2}{c}{Secondary} \\
Parameter & Value & $\pm$ & Value & $\pm$ \\
\hline
$i$ [$^\circ$] 	& \multicolumn{4}{c}{$82 \pm 2$}\\
$a$ [R$_\odot$]	& \multicolumn{4}{c}{$2.502 \pm 0.035$} \\
$M$ [M$_\odot]$	& 0.612 & 0.030 & 0.445 & 0.019 \\
$\Omega$	&  4.93 & 0.13  &  5.21 & 0.19  \\
$R$ [R$_\odot]$	& 0.596 & 0.020 & 0.445 & 0.024 \\
$\log{g}$	&  4.68 & 0.31  & 4.79  & 0.45  \\
$v_{rot}$ [km s$^{-1}$]$^a$& 67.3 & 2.5 & 50.5 & 2.8 \\
$V-I$ [mag]$^b$	&  1.84 & 0.14  &  2.17 & 0.14  \\
$T_{eff}$ [K]	&  3750 &  250  &  3085 &  300  \\
$M_{bol}$ [mag]	&  7.76 & 0.30  &  9.23 & 0.45  \\
$M_V$ [mag]	&  9.18 & 0.31  & 10.65 & 0.89  \\
$d$ [pc]	& \multicolumn{4}{c}{$39 \pm 6$}\\
 \multicolumn{5}{c}{\it Parameters of spots}\\
Longitude [rad]$^c$&  0.20 & 0.05  & 4.71 & 0.07 \\
Radius [rad]	&  0.7  &  0.1  &  0.3 & 0.08 \\
Contrast 	&  0.9  &  0.1  &  0.9 & 0.1  \\
\hline
\end{tabular}
\flushleft
$^a$ In case of a synchronous rotation\\
$^b$ Colour dereddened with $E(V-I)=0.02$ mag\\
$^c$ Counted counter-clockwise with 0 defined as the direction
toward the companion star\\
\end{table}

\begin{figure*}
\includegraphics[width=0.85\textwidth]{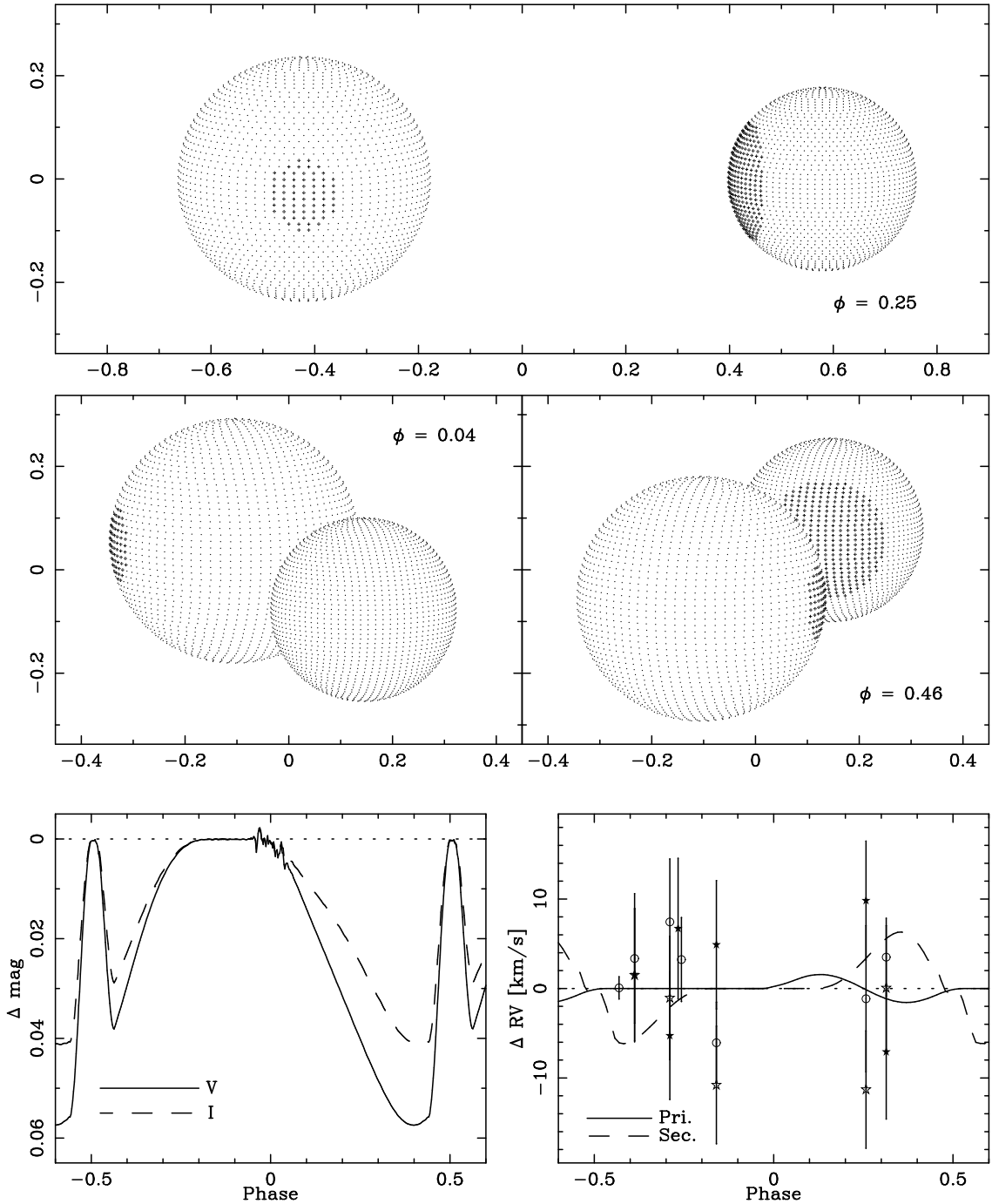}
\caption{{\it Top and middle:}
Three-dimensional reconstruction of the ASAS-01~A eclipsing binary, based on
our best-fitting model in quadrature (top) and during eclipses (middle). 
Dimentions are scaled so that $a = 1$. {\it Bottom:} Influence of
spots on the light curves (left) in $V$ (solid line) and $I$ (dashed) bands, and
on the RV curves of the primary (solid) and secondary (dashed) component. On the 
RV panel we also plot the residua of the model, in the same way as in Fig. 
\ref{fig_rv}. On the LC panel, the large scatter on the $V$ curve around 
$\phi\sim 0$ comes from the sampling of the stellar surface used in the \phoebe
code.
}\label{fig_spot}
\end{figure*}

The final absolute values of derived physical parameters of 
ASAS-01~A are collected in Table \ref{tab_par}. In Fig. \ref{fig_lc}
we present the observational and modeled light curves in $V$ (left) 
and $I$ (right) bands. One can see that the data taken with two 
different instruments and within almost a year overlap quite well,
so it was justified to use them simultaneously. As was mentioned
before, the pattern of the spots did not change significantly during 
this time, and no short-scale evolution occurred, as was found
for two other low-mass systems -- ASAS-09 and ASAS-21 \citep{hel11b}.
The 3D reconstruction of the eclipsing pair in three orbital phases 
-- 0.04 (primary eclipse), 0.25 (quadrature), and 0.46 (secondary
eclipse) -- is presented in Fig. \ref{fig_spot} (top and middle), 
together with the influence of spots on the light curve (bottom left) 
and radial velocities (bottom right). On the $\Delta$RV plot we also 
present RV residuals in the same form as in Fig. \ref{fig_rv}. 
One can see that the influence of spots on the RV curves is much 
smaller than the spread of measurements, which allows 
us to conclude that the uncertainty of the mass is not underestimated. 
We did not succeed in reaching a 3\% level of precision in mass and 
radius determination but we were close (3.4 -- 5.4\%). 

One can also note that our distance determination -- 39(6)~pc -- 
is in very good agreement with the result obtained by \citet{ria06}.
Assuming the angular separation of 1.405(3) asec \citep{ber10}, 
the projected physical separation of the A and B components is thus 
$\hat{a}_{AB}$~=~55(8)~AU. This corresponds to an orbital period 
$P_{AB} \simeq 333$ yr, assuming a circular orbit and the total mass 
of component B $M_B \simeq 0.45$ M$_\odot$ (similar to $M_{Ab}$
because of almost exactly the same value of $V-I$), or 292~yr
for $M_B \simeq 0.9$ M$_\odot$ (see Sect. \ref{sec_B} for a
discussion). The gravitational influence of B on the pair A would
not be detected with eclipse timing or precise RV measurements;
however the orbital motion of B should be detectable with adaptive
optics or various interferometric techniques.

\section{Discusion}
\subsection{Evolutionary status}
We compare our results with three sets of theoretical isochrones:
Yonsei-Yale \citep[Y$^2$;][]{yi01,dem04}, Dartmouth \citep{dot07}, 
and Padova \citep{gir00,mar08}. 

\begin{figure}
\centering
\includegraphics[width=0.79\columnwidth]{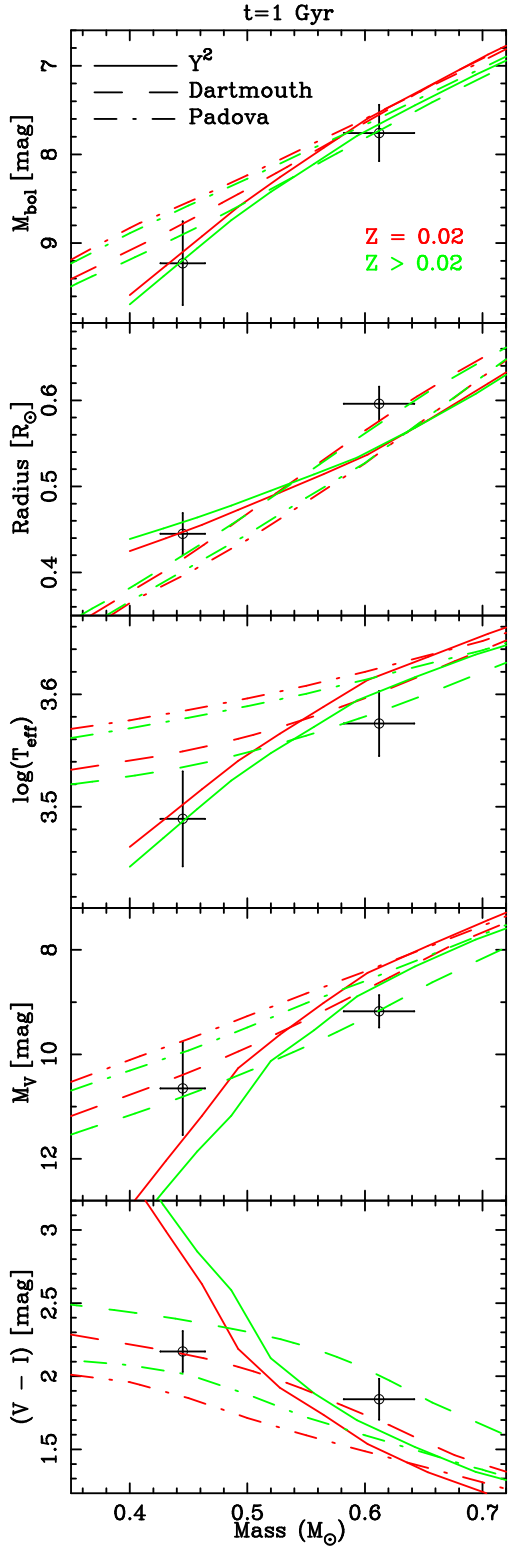}
\caption{Comparison of our results for ASAS-01~A eclipsing pair
with theoretical 1 gyr isochrones on mass $M$ vs. (from top to bottom):
bolometric magnitude, radius, effective temperature, absolute $V$ 
magnitude, and $V-I$ colour. \yy isochrones are depicted with solid, 
Darthmouth with dashed, and Padova with dot-dashed line. Isochrones for 
solar metallicity ($Z\simeq 0.02$) are depicted with red and 
for higher $Z$ with green. The colour version of the picture is 
available in the on-line version of the paper.}\label{fig_evo}
\end{figure}

The comparison is shown in Figure \ref{fig_evo}. We present 
the bolometric magnitude, radius, effective temperature, absolute $V$ 
magnitude and the $V-I$ colour as a function of stellar mass. We compare
our results with isochrones for an age of 1 Gyr and two cases of
metallicities: (1) solar, $Z\simeq0.02$ (red lines), and (2) above
solar (green lines), however different for every set: 
$Z=0.04$ for Y$^2$, 0.035 for Dartmouth, and 0.03 for Padova. 
Since the accuracy of our measurements is worse than the canonical 3\%, 
the isochrones are shown only for comparison, i.e. we do not attempt
to determine the age of the system. It is however
enough to conclude that ASAS-01 is probably a main sequence object.
The metallicity determination is even more insecure, however from
the Fig. \ref{fig_evo} one can claim that metallicities higher 
than solar are prefered -- on the $M$ vs. $M_{bol}$ and $\log{T_{eff}}$
planes the $Z>0.02$ isochrones fit significantly better to the data 
points.

The main sequence evolutionary stage of ASAS-01 is confirmed
by its galactic kinematics. We used our determinations of the systemic 
velocity and distance together with the position and proper motion 
from the PPMXL catalogue: $\mu_\alpha = 120.4\pm4.0$~mas~yr$^{-1}$,
$\mu_\delta = -36.7\pm4.0$~mas~yr$^{-1}$ \citep{roe10}. The obtained 
values of $U = 3.3\pm2.3$~km~s$^{-1}$, $V = -16.8\pm2.1$~km~s$^{-1}$ and
$W = -7.8\pm2.3$~km~s$^{-1}$ put ASAS-01 in the galactic thin disk,
and only marginally in one of the young moving groups recently reported
by \citet[][ID=15 therein]{zha09}.

It is worth noticing that on the $M/R$ plane the measurements 
agree within errors with the \yy and Dartmouth models. The second one
especially seems to reproduce the physical properties of 
ASAS-01~A components. It is in a somewhat contrary to many recent
results of low-mass detached eclipsing binaries studies
\citep[see:][]{kra11}. The usually, but not always, observed characteristic
of low-mass stars in close binaries is that theoretically predicted
radii are smaller and temperatures larger with respect to what is 
observed \citep{rib08}, and this situation is seen for the Padova 
models. The Dartmouth models also seem to follow that trend, but 
only by a few per cent. Within our uncertainties they correctly 
predict the observed radii, and, in the case of $Z=0.035$, also 
temperatures. The deviation is smaller for the larger component, 
which presumably rotates faster, thus it does not support the 
possible radius-rotation relation for $M<0.7$~M$_\odot$ stars
\citep{kra11}. At the same time the properties of the secondary 
are very well reproduced by the \yy models (except for the colour)
but for the primary we see the ``typical'' underestimation of
the radius and overestimation of the temperature. From this 
short discussion we can only conclude that the Dartmouth models 
may be the best to reproduce the properties of 
low-mass stars \citep[see also:][]{tho10} although
discrepancies at the level of $\sim$3\% are present. This is
however significantly smaller than for the majority of previous
studies where 5-15\% difference in radii were claimed.

\subsection{The companion B}\label{sec_B}

Having the two components separated on $V$ and $I$ band images
(Fig. \ref{fig_foto}), and the physical model created with
\phoebe with the third light included, we were able to estimate the
photometric properties of component B. We present the 
fractional fluxes in both bands, the dereddened $V-I$ colour, 
and the absolute $V$ magnitude in Table \ref{tab_3rd}.

\begin{table}
\centering
\caption{Photometric properties of the component B from the 
\phoebe model}\label{tab_3rd}
\begin{tabular}{lrl}
\hline \hline
Parameter & Value & $\pm$ \\
\hline
$F_{V,3}$ [\%]$^a$	& 34.8 & 1.5 \\
$F_{I,3}$ [\%]$^a$ 	& 42.9 & 3.2 \\
$(V-I)_3$ [mag]$^b$	& 2.15 & 0.13 \\
$M_{V,3}$ [mag]		& 10.82 & 0.70 \\
\hline
\end{tabular}
\flushleft
$^a$ Fractional fluxes defined as a percentage of the
flux of component A\\
$^b$ Colour dereddened with $E(V-I)=0.02$ mag
\end{table}

\begin{table}
\centering
\caption{Values of the inclination of the component B putative
orbit for various values of $q_B$ and masses of the more massive
component of 0.45 M$_\odot$ ($i_{B,0.45}$) and
0.5 M$_\odot$ ($i_{B,0.5}$). The mass function is taken 
from the Table \ref{tab_rv_B}.}\label{tab_q_3rd}
\begin{tabular}{ccccc}
\hline \hline
$q_B$ & $M_{Ba}\sin^3{i_B}$ & $M_{Bb}\sin^3{i_B}$ & $i_{B,0.45}$ & $i_{B,0.5}$\\
 & [M$_\odot$] & [M$_\odot$] & [$^\circ$] & [$^\circ$] \\
\hline
1.00	&0.226		&0.226  	 &53	  &50\\
0.95	&0.251		&0.238  	 &55	  &53\\
0.90	&0.280		&0.252  	 &59	  &56\\
0.85	&0.315		&0.268  	 &63	  &59\\
0.80	&0.358		&0.286  	 &68	  &63\\
... &&&& \\
0.718	&0.45		&0.323  	 &90	  &75\\
0.684	&0.5		&0.342  	 &---	  &90\\
\hline
\multicolumn{5}{l}{$f(M_{Ba}) = 0.0565(92)$ M$_\odot$} \\
\end{tabular}
\end{table}

\begin{figure}
\centering
\includegraphics[width=\columnwidth]{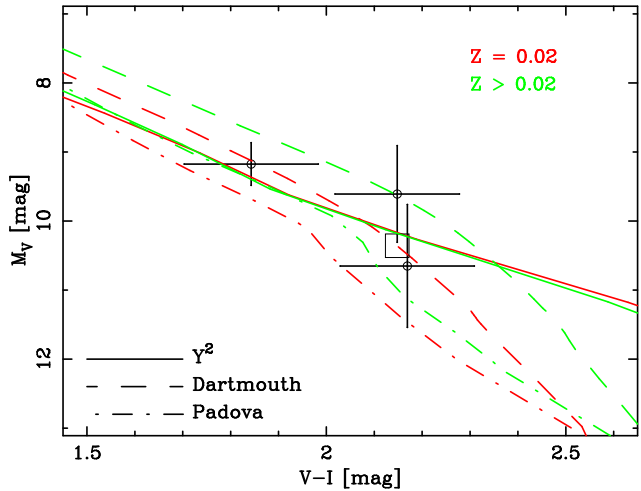}
\caption{Colour-brightness diagram with our measurements for
ASAS-01~Aa, Ab and B components, together with theoretical 
1~Gyr isochrones. Line and colour coding is the same as in Fig. 
\ref{fig_evo}. The square shows the approximate position of the components
of ASAS-01~B if it was composed of two identical stars showing the same
$V-I$ value as observed. The colour version of the Figure is 
available in the on-line version of the manuscript.}\label{fig_3rd}
\end{figure}

In Figure \ref{fig_3rd} we present our measurements on the 
$V~-~I/M_V$ plane. Data from Tables \ref{tab_par} and 
\ref{tab_3rd}, and the same isochrones as in Fig. \ref{fig_evo}
are used. We see that again the Dartmouth set for $Z=0.02$
gives the best match for the eclipsing pair A. Component
B is about 0.7~mag brighter but has almost the same colour 
as Ab ($V-I$ = 2.14 mag). Due to relatively large errorbars 
(magnified mainly by the uncertainty in the third light 
parameter in our model) one can formally find $V$ magnitudes 
of B and Ab almost equal and still rather consistent with the 
considered Dartmouth and both \yy isochrones. However, if assuming 
that component B is composed of two nearly-twin stars 
with masses $\sim$0.45-0.5~M$_\odot$, which would thus have 
the $V-I$ colour around 2.15~mag, i.e. almost the same as the
value observed for B, those putative Ba and Bb components
would be located in Fig. \ref{fig_3rd} on the position
marked by the empty square (lowering the flux by a factor of
2 increases the magnitude by $\sim$0.7). In such a situation,
the Darthmouth $Z=0.02$ and \yy isochrones would
reproduce the observed properties of the system much better. 
We consider this fact as a support for the hypothesis of 
the binarity of ASAS-01~B.

The binarity of component B, with a mass ratio close to
1 and component colours close to the value observed, 
is consistent with the fact that no marks of additional
eclipses are found in our $V$ and $I$ band light curves.
From the orbital fit for component B radial velocities,
presented in Table \ref{tab_rv_B}, we can estimate the 
orbital inclination $i_B$ for various values of the mass-ratio
$q_B$ and the desired mass of the more massive component $M_{Ba}$.
We present those calculations in Table \ref{tab_q_3rd}.

For a number of different values of $q_B$, we present 
values of $M \sin^3{i_B}$ for Ba and Bb, and values of 
inclination angles, for which $M_{Ba} = 0.45$ or 0.5 M$_\odot$.
From this Table one can see, that if ASAS-01~B is a binary,
to explain all the observed properties one needs the inclination
angle $i_B$ to be between 50 and 70 degrees, which corresponds
to $q\ge 0.8$. For these values of masses of Ba and Bb we can expect 
the observed colour to be around the value given in 
Tab. \ref{tab_3rd}. Assuming lower values of $q_B$ we would end 
up in a situation where additional eclipses occur and the
Bb component is significantly fainter. In such a
case the total brightness of ASAS-01~B would be lower than it is
observed, however still within relatively large error bars.

If ASAS-01 is a double-double system, it would join a small group 
of this kind of interesting objects. To date only two other
low-mass detached eclipsing binaries are known to be in similar
configuration: BD~-22~5866 -- a system with K7+K7 eclipsing pair 
and M1+M2 non-eclipsing binary \citep{shk08}, and YY~Gem -- the faintest 
member of a sextuple system composed of three spectroscopic binaries 
-- Castor A, B and C \citep[$\alpha$~Gem~ABC;][]{vin40,kro52,bop74}. 
There are also few examples of LMDEBs known to have a single
additional companion, like LP~133-373 \citep{vac07}, 
HIP~96515\footnote{No RV curve is published for this system.}
\citep{hue09}, MR~Del \citep{pri09,dju11}, NLTT~41135 \citep{irw10}, 
ASAS-08 \citep{mon07,hel11a} or the triply-eclipsing KOI-126 \citep{car11}. 
Such systems not only allow more 
rigid constrains on the evolutionary models than in the cases of ``lonely'' 
eclipsing binaries, but also play important role in testing star-formation
theories, stellar population codes and dynamical interactions
in multiple stellar systems. The relative brightness and
small distance to ASAS-01 make it a valuable object for 
further studies.

\section{Summary}
As a result of our analysis of the components of ASAS~J01328-3821.1, 
an M-type multiple system, we obtained a complete set of 
orbital and physical parameters of the detached eclipsing pair A, 
and photometric and spectroscopic properties of component B.
Due to a number of difficulties our results are of an accuracy of 3-6 \%,
which is not enough for performing reliable estimation of the 
evolutionary status of the system. Within the errorbars we may
however reproduce our results with main sequence theoretical models.
The observed spectroscopic and photometric
properties of component B of the system
suggest that this is a binary composed of nearly identical
stars. However the eventual orbital solution is quite uncertain,
and $M_V$ and $V-I$ errors are quite large, thus we can not make
any final conclusions on this topic. ASAS-01, as a member of a rare
class of late-type multiples with eclipsing detached components, 
should be considered as a potential testbed for stellar formation
and evolution theories and definitely deserves more 
attention and investigation, especially considering the system's 
spectroscopy and photometric properties of component B.

\
\section*{Acknowledgments}

We would like to thank David Laney, John Menzies and Hannah Worters from the 
South African Astronomical Observatory for their support during our observations
at SAAO, and Stephen Marsden and the Anglo-Australian Observatory 
astronomers for their help during our observing runs on the AAT.
We thank Samba Fall and Chelsea Harrison for careful 
proofreading and corrections of this manuscript.

This research was co-financed by the European Social Fund and the national
budget of the Republic of Poland within the framework of the Integrated
Regional Operational Programme, Measure 2.6. Regional innovation
strategies and transfer of knowledge - an individual project of the
Kuyavian-Pomeranian Voivodship ``Scholarships for Ph.D. students
2008/2009 - IROP''. This work is supported by by the Polish 
National Science Center grant 5813/B/H03/2011/40, by the Foundation for Polish Science 
through a FOCUS grant and fellowship and by the Polish Ministry of Science and 
Higher Education through grants N203 005 32/0449, N203 3020 35, and N203 379936. 
K.G.H. acknowledges support provided by the Proyecto FONDECYT Postdoctoral No. 3120153, 
the Centro de Astrof\'{i}sica FONDAP 
Proyecto 15010003, Comit\'{e} Mixto ESO-Chile and by the Ministry for the Economy,
Development, and Tourism's Programa Iniciativa Cient\'{i}fica Milenio
through grant P07-021-F, awarded to The Milky Way Millennium Nucleus.
M.W.M. acknowledges support from the Townes Fellowship Program, an internal UC Berkeley 
SSL grant, and the State of Tennessee Centers of Excellence program. 
This research was supported in part 
by the National Science Foundation under Grant No PHY05-51164 and through 
Grants 0959447, 0836187, 0707634 and 0449001. The observations on the
AAT/UCLES have been funded by the Optical Infrared Coordination network (OPTICON),
a major international collaboration supported by the Research Infrastructures 
Programme of the European Commissions Sixth Framework Programme.

This research has made use of the Simbad database, operated at CDS, Strasbourg, France. 

\appendix

\bsp

\label{lastpage}

\end{document}